\documentclass[aps,prl,preprint,superscriptaddress]{revtex4}
\usepackage{graphicx}% Include figure files
\usepackage{dcolumn}% Align table columns on decimal point
\usepackage{bm}% bold math
\begin{document}

%\documentstyle[epsfig,preprint,prl,aps]{revtex}
%\begin{document}

%\draft

\title{A study of the $\bf \Delta^-$-component of the wave-function in
light nuclei.}

\author{E. A. Pasyuk} \affiliation{Joint Institute for Nuclear Research,Dubna, Moscow Region, 141980 Russia}
\altaffiliation[Present address:] {Physics Department, Arizona State University, Tempe, AZ 85287}
\author{R. L. Boudrie}\affiliation{Los Alamos National Laboratory, Los Alamos, NM 87545, USA}
\author{P. A. M. Gram}\affiliation{Los Alamos National Laboratory, Los Alamos, NM 87545, USA}
\altaffiliation[Present address:] {P.\ O.\ Box 173, Captain Cook, HI 96704}
\author{C. L. Morris}\affiliation{Los Alamos National Laboratory, Los Alamos, NM 87545, USA}
\author{J. D. Zumbro}\affiliation{Los Alamos National Laboratory, Los Alamos, NM 87545, USA}
\author{J. L. Matthews}\affiliation{Massachusetts Institute of Technology, Cambridge, MA 02139, USA}
\author{Y. Tan}\affiliation{Massachusetts Institute of Technology, Cambridge, MA 02139, USA}
\author{V. V. Zelevinsky}\affiliation{Massachusetts Institute of Technology, Cambridge, MA 02139, USA}
\altaffiliation[Present address:] {1822 Beacon St.\, \#3, Brookline, MA 02445}
\author{ G. Glass}\affiliation{3218C Walnut St., Los Alamos, NM 87544, USA}

\author{B. J. Kriss}\affiliation{University of Colorado, Boulder, CO 80309}
\altaffiliation[Present address:] {Lockheed Martin Mission Systems, Colorado Springs, CO 80921}
\date{\today}
\begin{abstract}
We have measured cross sections for the $ (\pi^+,\pi^\pm p)$ 
reactions on ${\rm ^3H}$, ${\rm ^4He}$, ${\rm ^6Li}$ and ${\rm ^7Li}$ in  
quasi-free kinematics at incident pion beam energy 500~MeV. 
An enhancement of the $(\pi^+,\pi^- p)$ cross section in this kinematics is observed.  
If this is interpreted as due to quasi-free scattering from pre-existing 
$\Delta$ components of the nuclear wave function, 
the extracted probabilities are in agreement with theoretical expectations.
\end{abstract}
\maketitle

Although in conventional models the atomic nucleus is made up of protons, 
neutrons and virtual pions, there is an intriguing possibility that other 
particles exist with small probabilities.  
If this is the case, cross sections for some reactions can be dramatically 
enhanced because of new degrees of freedom introduced by these virtual particles. 
Theoretical investigations \cite{Pri69,Ker69} in the 1960's first suggested that 
nucleon resonances might play a role in nuclear structure.  
The lowest excited state of the nucleon, the $\Delta$, is expected to be 
particularly important because of its low excitation energy (300 MeV) and 
its strong coupling to the pion-nucleon system.  
Indeed, meson models of nuclear binding predict virtual excitation of $\Delta$s 
in finite nuclei with probabilities of a few percent.\cite{Fri83} 
Detecting virtual $\Delta$'s at these levels has proven difficult, 
since $\Delta$ production reactions usually dominate over $\Delta$ knockout 
reactions. \cite{Tat78,All86}

Due to isospin selection rules, 
the lightest nuclei in which single-$\Delta$ wave-function components can occur 
are $^3$H and $^3$He.  Because of the existence of extensive 
calculations, \cite{Str87,Haj83a,Haj83b,Pen93,Pic91a,Pic92a,Pic92b,Pic92c,Pic95} 
mass-3 nuclei provide the ideal testing ground for comparing theory with experiment. 
The $\Delta$ probabilities predicted in this body of work range from 1\% to 3\%. 

Pion electroproduction has been suggested to be an ideal probe to search for 
preformed $\Delta$'s in $^3$He because of the possibility of studying the longitudinal 
channel, in which $\Delta$ knockout is enhanced over $\Delta$ production, compared 
with the transverse channel. 
In addition, the Coulomb interaction favors $\Delta$ knockout in $^3$He since 
isospin couplings enhance the $\Delta^{++}$ fraction in this nucleus. \cite{Lip87,Mil88} 
However, experimental searches for measurable consequences of $\Delta$ components in 
the three-nucleon system using electroproduction have been 
inconclusive. \cite{Adb90,Emu93,Blo96}

Recently, a new method for measuring the $\Delta^-$ component of the nuclear wave 
function has been suggested and has yielded results which are in agreement with the 
predictions of meson models of nuclear binding for range of nuclei. \cite{Mor96,Mor98} 
This method is based on pion double charge exchange (DCX) which cannot occur in a 
single step if the pion interacts with nucleons. However, since the isospin-3/2 $\Delta$ 
exists in four charge states, the double-charge-exchange reaction $\Delta^-(\pi^+,\pi^-)p$ 
can occur in a single step. 
This leads to the expectation of a significant enhancement in the cross section for DCX 
on $\Delta$ components of the nuclear wave function in quasielastic kinematics 
with respect to the two-step background. 
Further, the cross section for this process is expected to be large 
because there is no need to transfer 300~MeV to scatter the $\Delta$ on-shell 
as in the electroproduction experiments.  
In the current work, we extend the study of  ($\pi^+,\pi^-p$), 
to the lightest stable nucleus for which this reaction is possible, $^3$H. 
Measurements were also made on ${\rm ^4He}$, ${\rm ^6Li}$, and ${\rm ^7Li}$.

The experiment was performed using a 500~MeV pion beam from the P$^3$-East channel 
at the Clinton~P.~Anderson Meson Physics Facility. Scattered pions from both the 
($\pi^+,\pi^+p$) [NCX] and ($\pi^+,\pi^-p$) [DCX] reactions were observed using the 
LAS spectrometer\cite{LAS} at an angle of 50$^{\circ}$. This gives a momentum transfer, 
$q$, of about 488 MeV/{\it c}. The momenta of coincident protons were measured using a 
second magnetic spectrometer (BAS) at 52$^{\circ}$ on the opposite side of the beam, near 
the direction of the recoil proton in free $\pi{p}$ scattering. 
Both spectrometers measured the particle positions and angles before and after a bend. 
The particle trajectory from the target and the momentum were reconstructed using this 
information. Helium bags were used in the particle paths in both spectrometers to minimize 
multiple scattering. The BAS spectrometer consisted of a single rectangular dipole arranged 
to provide a bend of 30$^{\circ}$ for the central momentum. 
The
momentum acceptance was approximately 20\%, and the solid angle was 7 msr. 
The momentum resolution, limited by the angle measurement, was 1\%.

The angular acceptance of both spectrometers spanned a full width of about 10$^{\circ}$ and was trapezoidal in shape. This corresponds to a transverse momentum acceptance
Of $\pm$50 MeV/c at the quasielastic peak, only a small fraction of the quasielastic phase space.

The $^3$H target was contained in a 3.5~cm radius high strength stainless steel 
sphere at a pressure of 102.7~bar. The $^4$He target was contained in a conventional 
20~cm diameter aluminum wall gas cylinder at a pressure of 134.6~bar.  
The interaction vertex was reconstructed using the trajectory information to remove 
the background from gas cell walls. The ${\rm ^{6,7}Li}$ targets consisted of plane 
foils of thickness 0.616 and 0.576~g/cm$^2,$ respectively.  

\begin{figure}
\includegraphics [width=4.25in,height=5.5in]{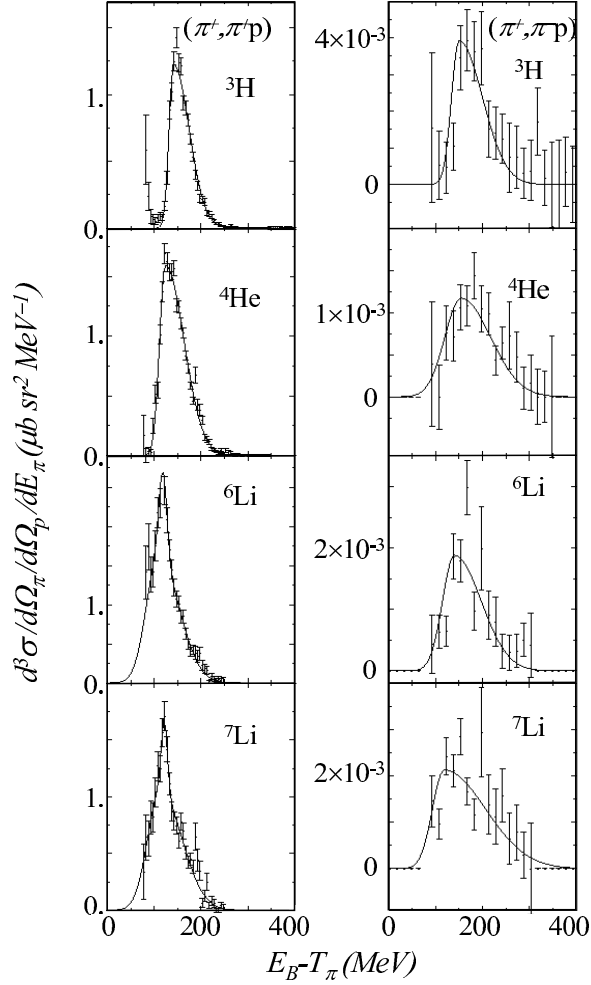}
\caption{\label{alldata} Spectra of outgoing pion energy loss for: NCX (left), and DCX
 (right). Solid lines display the results of fits to obtain the widths.}
\end{figure}

\begin{figure}
\includegraphics [width=6in,height=4.5in]{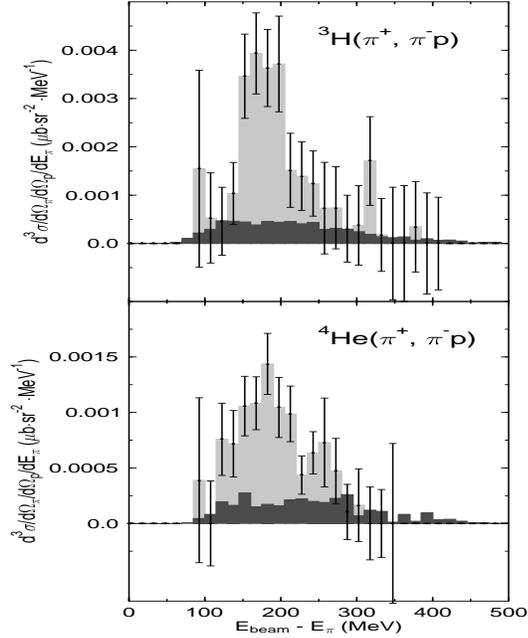}
\caption{\label{34dcx}
Spectra of outgoing pion energy loss for: $^3$H ${\rm (\pi^+,\pi^\pm p)}$ (top), 
and $^3$H ${\rm (\pi^+,\pi^\pm p)}$ (bottom). 
The dark shaded region is a phase-space calculation discussed in the text. 
The light shaded region emphasizes the data.}
\end{figure}

A clear signature of coincident protons was obtained by measuring the time differences 
between the pion and the proton and by using the energy-loss signals from scintillation 
detectors for both particles. Backgrounds due to random coincidences and other 
processes were found to be very small, in the DCX as well as in the NCX spectra. 
Even though the NCX cross sections are a factor of $\approx$1000 larger than the 
DCX cross sections, the use of a magnetic spectrometer provides a potent filter 
against pions of incorrect charge. In the current analysis the sum of the pion 
and proton energies was required to exceed 400~MeV ({\it i.e.}, the missing energy 
was required to be less than 100 MeV).  
Pion energy loss spectra are presented in Fig.~\ref{alldata} for both the NCX 
(left panel) and DCX (right panel) reactions. 
Figure~\ref{34dcx} displays the DCX data for $^3$H and $^4$He along with phase 
space for the DCX reaction, which is a 4-body final state in the case of $^3$H and a 
5-body final state for $^4$He, shown as dark shaded histograms. The phase space 
calculations have been normalized to the average of the data above 250 MeV of energy loss.  
The DCX data clearly exhibit a peak over and above $n$-body phase space, consistent 
with the results of the previous experiment.\cite{Mor96,Mor98}
In Ref.~\cite{Mor98}, the DCX data were compared with the prediction of an 
intranuclear cascade model,\cite{Gibbs} in which $(\pi^+, \pi^- p)$ was assumed to 
result from the sequential processes 
$\pi^+ n \rightarrow \pi^0 p$ and $\pi^0 n \rightarrow \pi^- p$.  
The calculation provided a good representation of the NCX data but underestimated 
the DCX data in the region of the peak by more than a factor of five.\cite{Mor98}

The outgoing pion energy-loss spectra were fitted with asymmetric Gaussian 
lineshapes to obtain the full widths at half maximum of the peaks, $\Gamma$. 
In the case of the NCX spectra on $^6$Li and $^7$Li two Gaussians were 
required to fit both a narrow component of the quasi-free, QF, 
peak presumably due to knockout from the $p$-shell, and a wide component due to 
knockout from the $s$-shell.  The solid lines shown in Fig.~\ref{alldata} 
display the fits. Fermi momenta, $k_F$ ($k_{\Delta}$), for the knocked-out 
proton ($\Delta$) were calculated using the non-relativistic 
expression \cite{Peterson-92}:
\begin{equation}
k = {{\Gamma M \sqrt{2}}\over{q}},\label{q}
\end{equation}
where $M$ is the mass of the knocked out particle. 
We have assumed this to be the proton mass for both  NCX and DCX. 
The resulting widths and momenta are given in Table~\ref{results}.

The significantly smaller value of $k_{\Delta}$ for $^3$H relative to 
that for $^4$He can be qualitatively understood as arising from the allowed 
spin couplings for the $\Delta^{-}$ in the two nuclei.  
The process by which the $\Delta^{-}$ is created is 
$n + n \rightarrow p + \Delta^{-}$. The $p$ and $\Delta^{-}$ 
spins can couple to 1 or 2, which must couple to the spin of the 
spectator nucleon(s) to yield the ground-state spin of the nucleus. 
In $^3$He, with one spectator proton, this is clearly possible. 
In $^4$He, however, the two protons in the dominant 1$s_1/2$ state must have 
total spin 0, and thus the $p + \Delta^{-}$ can only couple to core excited 
configurations of the ground state, which are expected to have higher momentum 
components.

An estimate of the probabilities for pre-existing $\Delta$'s can be obtained 
from the data by integrating the cross sections and constructing the ratio, $R$, 
of DCX to NCX cross sections. This cancels distortion factors and leaves
\begin{equation}
P_{\Delta} = R \left({{ZN_{\Delta}}\over{AN_{\Delta^-}}}\right)
\left({{\sigma(\pi^+ + p \rightarrow \pi^+ + p)} \over
{\sigma(\pi^+ + \Delta^- \rightarrow \pi^- + p)}}\right)
\left({{k_{\Delta}}\over{k_F}}\right)^2,\label{PD}
\end{equation}
where $P_{\Delta}$ is the sum of the probabilities for all $\Delta$ charge states, 
$N_{\Delta^-}/N_{\Delta}$ is the number of $\Delta^-$'s over the sum of $\Delta$'s 
in all charge states, $Z$ and $A$ are the proton and nucleon numbers and account 
for the number of protons (measured in NCX) per nucleon, $k_{\Delta}$ characterizes 
the momentum spread of the $\Delta$ component of the wave function, and $k_F$ is 
the momentum spread of the nucleons. We have used measured branching ratios \cite{PPDB} 
for $\rm{N}^* (1520) \rightarrow \pi + \rm{N}$ and 
$\rm{N}^* (1520) \rightarrow \pi + \Delta(1232)$, 
corrected for two-body phase space with the appropriate isospin Clebsch-Gordan 
coefficients, to estimate the cross-section ratio in Eq.~(\ref{PD}) to be 1.46$\pm$0.32.
The uncertainty in the cross section ratio results in a scaling error that applies to all of the delta probabilities 
reported in this work and has not been included in the error estimates. 
The increased momentum spread decreases the acceptance for $\pi$-p coincidences as 
$(k_{\Delta}/k_F)^2$ due to the small fraction of quasielastic phase covered in this experiment.  These values have been obtained from fitting the data.
The ratios of $\Delta$ charge states have been obtained by generalizing arguments 
given in Ref.~\cite{Lip87} to give
\begin{equation}
{{N_{\Delta^-}}\over{N_{\Delta}}} = {{3}\over{4}} \left(1 +
{{Z}\over{2(N-1)}} + {{Z(Z-1)}\over{N(N-1)}}\right)^{-1},\label{RD}
\end{equation}
where $N$ is the neutron number of the nucleus. 
Typographical errors in ref. \cite{Mor98} have been corrected in the above two equations.
The relative probabilities for each of the $\Delta$ charge states are assumed to be given 
by sums of isospin Clebsch--Gordan coefficients multiplied by the appropriate number 
of T=1 paired nucleons: $N(N-1)$, $NZ/2$, and $Z(Z-1)$ for proton-proton, neutron-proton, 
and proton-proton pairs, respectively.

The measured cross section ratios, $R = \sigma_{DCX}/\sigma_{NCX}$, 
and resulting $\Delta$ probabilities for all targets are given in Table~\ref{results}.
Figure~\ref{adepn} shows the A dependence of the extracted
$\Delta$ fraction as closed symbols along with previous data (open symbols) taken at 
pion angles of 
50${^\circ}$. \cite{Mor98}  
In addition to the statistical error given in the table there is an additional 
30\% error introduced by uncertainties in the estimates of the momentum distributions.
\begingroup
%\squeezetable
\begin{table}
\caption{Measured widths of NCX and DCX peaks, extracted Fermi momenta of the pions 
and $\Delta$'s, measured cross section ratios, and extracted values of
$P_{\Delta^-}$ and $P_{\Delta}$ the $\Delta^-$ and $\Delta$ probabilities
respectively. In addition to the statistical uncertainty given 
in the table there is an additional 30\% uncertainty in the $\Delta$ probabilities 
introduced by uncertainties in the estimates of the momentum distributions.}
\label{results}
\begin{tabular}{cccccccccc}
Target &$\Gamma_{NCX}$&$\Gamma_{DCX}$&$k_{Fermi}$&$k_{\Delta}$&
$\sigma_{\rm{DCX}}/\sigma_{\rm{NCX}}$ & $P_{\Delta^-}$ & $ P_{\Delta}$ \\
 &MeV&MeV&MeV/c&MeV/c&($\times 10^3$) & ($\times 10^3$)& ($\times 10^2$) \\
\colrule
$^3$H & 53.0(1.1)& 77(13)& 151(3)& 218(38)& 6.0(8)   & 6.1(8) & 1.23(16) \\
$^4$He& 64.3(1.0)& 119(22)& 183(3)& 338(63)& 1.52(16) & 3.8(4)  & 1.52(16) \\
$^6$Li& 93.8(4.3)& 98(16)& 267(12)& 279(46)& 1.82(18) & 1.45(14)  & 0.53(5) \\
$^7$Li& 95.6(4.0)& 171(34)& 272(11)& 486(97)& 4.0(4)   & 7.9(7)  & 2.11(20) \\
\end{tabular}
\end{table}
%\endgroup
\begin{figure}
\includegraphics [width=6in,height=4.5in]{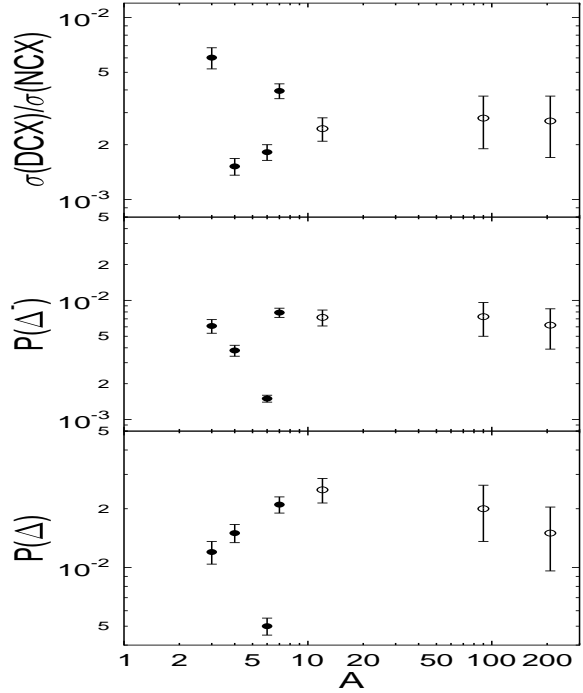}
\caption{\label{adepn} A-dependence of cross section ratios (top), extracted
$\Delta^-$ fraction (middle), and $\Delta$ fraction summed over charge states
(bottom). Data from the current experiment are shown as solid circles,
previous data \protect\cite{Mor98} are shown as open circles.}
\end{figure}
A difference of a factor of four is observed in the cross
sections and a factor of about two in the extracted $\Delta^-$ fractions 
between $^3$H and $^4$He. 
A factor of two results from the isospin Clebsch-Gordan coefficients 
favoring population of the $\Delta^-$ charge state in $^3$H. \cite{Lip87}  
A similar, larger, effect is observed in the comparison of the $\Delta^-$ 
fractions in $^6$Li and $^7$Li. The small $\Delta$ probability in $^6$Li may be 
due to the fact that the ground state of this nucleus like $^2$H, has isospin 0.

This analysis yields two important conclusions. 
First, the $\Delta^-$ probability in ${\rm ^3H}$ 
appears to agree with theoretical expectations based on a meson model of 
nuclear structure. \cite{Str87,Haj83a,Haj83b,Pen93,Pic91a,Pic92a,Pic92b,Pic92c,Pic95}
Second, the $\Delta^-$ probabilities in the light self-conjugate nuclei are 
significantly smaller than the $\Delta^-$ probabilities in the T=1/2 nuclei. 
This suggests that the quantum numbers of excited nucleon degrees of freedom
are robust and are reflected in the allowed nuclear configurations of the  $\Delta^-$ admixtures.  
Both of these observations support the long standing hypothesis that excited nucleon 
degrees of freedom play a role in nuclear structure. \cite{Pri69,Ker69} 

This work has been supported in part by the United States Department of Energy.
The authors would like to thank R.\ J.\ Peterson for suggesting the analysis 
of the widths to extract Fermi momenta, C.\ Fred Moore for running shifts on the 
experiment, and Dharam Ahluwalia for carefully reading the manuscript.

\end{document}